\documentclass{article}

\usepackage[T1]{fontenc}
\usepackage[utf8]{inputenc}
\usepackage{tgpagella}
\usepackage{microtype}

\usepackage[margin=1in]{geometry}

\usepackage{amsmath,amssymb,amsfonts,mathtools,amsthm}

\newcommand{\negl}{\operatorname{negl}}

\usepackage{booktabs}
\usepackage{array}
\usepackage{makecell}
\usepackage{multirow}
\usepackage{longtable}
\usepackage{tabularx}
\usepackage{tabulary}
\usepackage{supertabular}
\usepackage{enumitem}
\usepackage{tablefootnote}

\usepackage{graphicx}
\usepackage{float}
\usepackage{wrapfig}
\usepackage[font=small,labelfont=bf]{caption} 
\usepackage{subcaption}                        

\usepackage{xcolor}
\usepackage{xspace}
\usepackage{textcomp}
\usepackage{pifont}

\usepackage{scrextend}

\usepackage{algorithm}
\usepackage{algpseudocode} 
\makeatletter
\newcommand{\theHalgorithm}{\arabic{algorithm}}               
\@ifundefined{ALG@line}{}{%
  \providecommand{\theHALG@line}{\theHalgorithm.\arabic{ALG@line}} 
}
\makeatother

\usepackage{tikz}
\usetikzlibrary{patterns}
\usepackage{pgfplots}
\usepgfplotslibrary{groupplots}
\pgfplotsset{compat=1.18} 

\usepackage{nicefrac}
\usepackage{changepage}
\usepackage{xargs}
\usepackage{lipsum}
\usepackage{endnotes}
\usepackage{siunitx}
\usepackage{fontawesome5}
\usepackage{setspace}
\usepackage{lscape}

\usepackage[round,semicolon]{natbib}

\definecolor{mydarkblue}{rgb}{0,0.08,0.45}
\usepackage[colorlinks,citecolor=mydarkblue,urlcolor=mydarkblue,linkcolor=mydarkblue]{hyperref}
\usepackage[capitalize,noabbrev]{cleveref}
\crefname{section}{Section}{Sections}
\Crefname{section}{Section}{Sections}
\crefname{table}{Table}{Tables}
\crefname{figure}{Figure}{Figures}
\crefname{algorithm}{Algorithm}{Algorithms}
\crefname{equation}{Eq.}{Eqs.}
\crefname{appendix}{Appendix}{Appendices}

\theoremstyle{plain}
\newtheorem{theorem}{Theorem}[section]

\theoremstyle{definition}

\theoremstyle{remark}

\usepackage{minitoc}

\newcommand{\method}{\texttt{EFU}}

\title{\textbf{EFU: Enforcing Federated Unlearning via Functional Encryption}\thanks{This paper has been accepted to CIKM 2025. This version is a preprint and not the officially published version.}}

\author{
\textbf{Samaneh Mohammadi\textsuperscript{$^{\dagger}$,1,2},
Vasileios Tsouvalas\textsuperscript{$^{\dagger}$,3},
Iraklis Symeonidis\textsuperscript{1},
Ali Balador\textsuperscript{2}}\\
\textbf{Tanir Ozcelebi\textsuperscript{3},
Francesco Flammini\textsuperscript{2}} and 
\textbf{Nirvana Meratnia\textsuperscript{3}}\\[1em]
\textsuperscript{1} RISE Research Institutes of Sweden \quad
\textsuperscript{2} Mälardalen University\\
\textsuperscript{3} Eindhoven University of Technology
}

\date{}

\begin{document}

\maketitle
\begingroup
\renewcommand\thefootnote{}\footnotetext{${}^{\dagger}$Equal contribution. Corresponding authors: samaneh.mohammadi@ri.se, v.tsouvalas@tue.nl}
\endgroup

\begin{abstract}
    Federated unlearning (FU) algorithms allow clients in federated settings to exercise their ``\emph{right to be forgotten}’’ by removing the influence of their data from a collaboratively trained model. Existing FU methods maintain data privacy by performing unlearning locally on the client-side and sending targeted updates to the server without exposing forgotten data; yet they often rely on server-side cooperation, revealing the client's intent and identity without enforcement guarantees — compromising autonomy and unlearning privacy. In this work, we propose \textbf{\method} (\underline{\textbf{E}}nforced \underline{\textbf{F}}ederated \underline{\textbf{U}}nlearning), a cryptographically enforced FU framework that enables clients to initiate unlearning while concealing its occurrence from the server. Specifically,~\method~leverages functional encryption to bind encrypted updates to specific aggregation functions, ensuring the server can neither perform unauthorized computations nor detect or skip unlearning requests. To further mask behavioral and parameter shifts in the aggregated model, we incorporate auxiliary unlearning losses based on adversarial examples and parameter importance regularization. Extensive experiments show that~\method~achieves near-random accuracy on forgotten data while maintaining performance comparable to full retraining across datasets and neural architectures — all while concealing unlearning intent from the server. Furthermore, we demonstrate that~\method~is agnostic to the underlying unlearning algorithm, enabling secure, function-hiding, and verifiable unlearning for any client-side FU mechanism that issues targeted updates.
\end{abstract}

\section{Introduction}

Federated learning (FL) has emerged as a powerful paradigm for collaboratively training models across multiple clients without sharing raw data — addressing privacy concerns while still benefiting from the collective intelligence of distributed data sources. However, recent privacy regulations — such as the European Union's General Data Protection Regulation (GDPR)~\cite{gdpr} and the California Consumer Privacy Act (CCPA)~\cite{ccpa} — mandate the ``\emph{right to be forgotten}'', requiring that both a user’s data and its influence be removed from any trained Machine Learning (ML) model~\cite{voigt2017eu}. In response, \emph{Machine Unlearning} (MU) was introduced to eliminate the impact of specific data in centralized settings, and subsequently, its federated counterpart — \emph{Federated Unlearning} (FU) — extends this capability to FL, allowing clients to revoke their prior contributions

Recent efforts in FU have introduced various mechanisms to remove the influence of specific data from collaboratively trained models. These approaches can be broadly classified into (i)~\emph{passive} strategies, where clients depart after issuing deletion requests and the server performs unlearning independently — typically via model rollback~\cite{liu2021federaser, fast}, and (ii)~\emph{active} strategies, where clients remain involved by submitting tailored updates~\cite{halimi2022federated, zhong2025unlearning, pan2025ga}. While passive methods rely on stored gradients or checkpoints — incurring significant computational and storage costs — active methods use gradient perturbation~\cite{pan2025ga, halimi2022federated} or update sparsification~\cite{zhong2025unlearning} to induce finer-grained forgetting by modifying the model's predictive behavior. Beyond unlearning itself, recent work has also addressed its \emph{verification}~\cite{verifi}, enabling users to certify that their data has been erased — thus supporting the ``\emph{right to verify}'' alongside the ``\emph{right to be forgotten}''.

Nonetheless, existing FU methods primarily address \emph{how to forget data} from trained models, but offer neither a mechanism for clients to \emph{enforce unlearning} nor guarantee that their \emph{requests remain hidden}. These approaches typically rely on server-side cooperation — such as executing model rollback or aggregating tailored updates — which inherently exposes the occurrence of unlearning to the server. As a result, the client's intent to unlearn becomes explicitly observable and potentially traceable, enabling the server to not only identify such requests but also to omit, delay, or disregard them — undermining both the confidentiality and enforceability of unlearning. Achieving private, autonomous, and verifiable unlearning in realistic FL settings thus requires both \emph{function hiding} and \emph{enforcement} — that is, the ability to conceal whether an update corresponds to learning or unlearning, and to ensure its execution on the server-side.

\begin{figure*}[!t]
    \centering
    \includegraphics[width=0.9\linewidth]{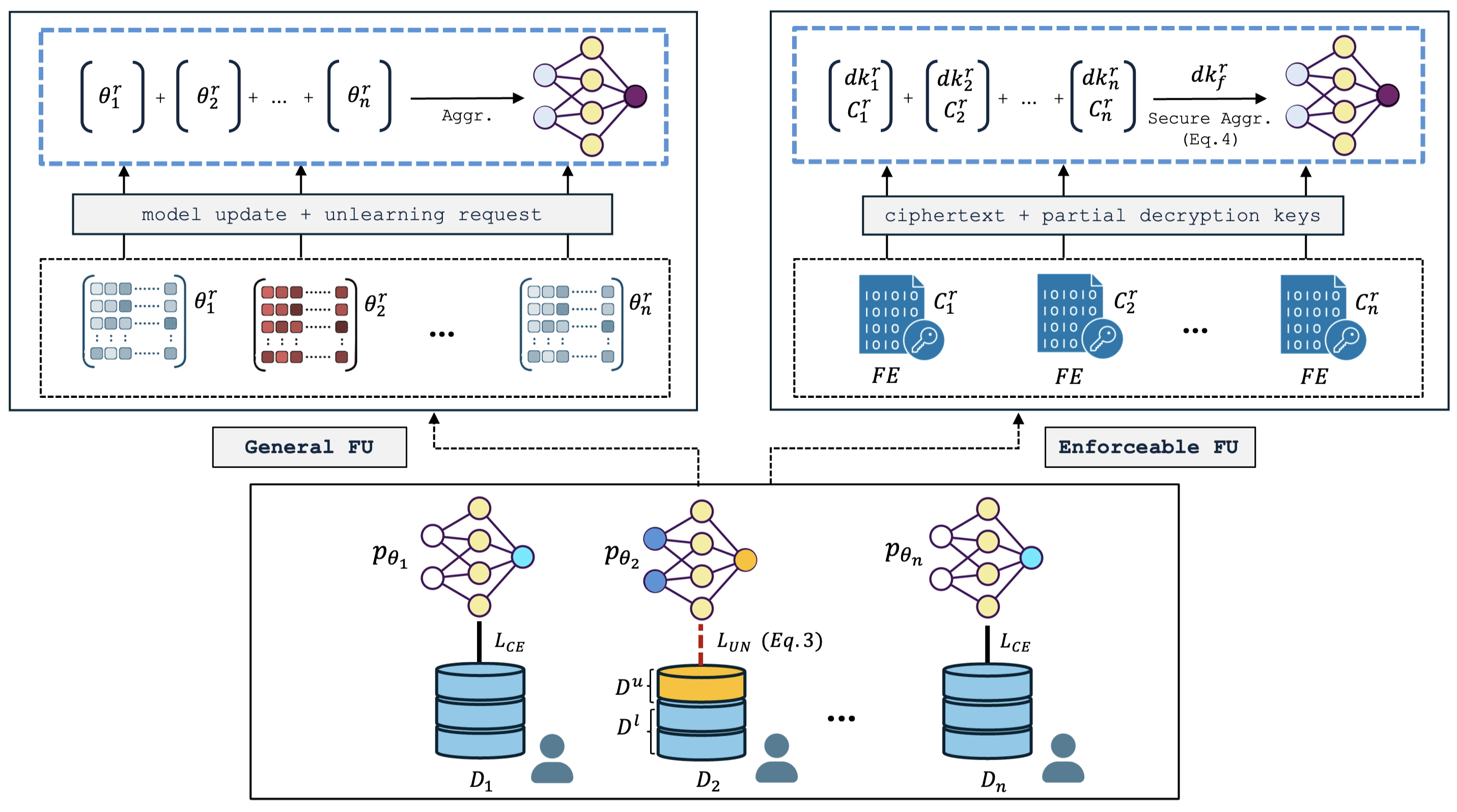}
    \caption{Overview of Federated Unlearning (FU) and \emph{E}nforceable \emph{F}ederated \emph{U}nlearning (\method). In standard FU, plaintext model updates \( \theta^r_i \) from both learning and unlearning clients reveal their intent and depend on server (\textcolor{blue}{blue box}) cooperation.~\method~encrypts model updates via functional encryption, binding them to a fixed aggregation that \emph{enforces} unlearning at decryption and \emph{conceals} client intent — without server trust.~\method~is a \emph{drop-in solution} compatible with any FU method that uses unlearning updates, requiring no changes to model architecture or training routines.}
    \label{fig:overview}
\end{figure*}

To fill this gap, we introduce~\method~(\underline{\textbf{E}}nforced \underline{\textbf{F}}ederated \underline{\textbf{U}}nlearn-ing), an FU approach that enables clients to initiate and cryptographically enforce the removal of their data influence from the federated model.~\method~leverages Functional Encryption (FE) to encrypt and bind each client update — whether for learning or unlearning — to a fixed aggregation function, ensuring that the server can neither alter the operation nor distinguish between update types. As a result, the unlearning process is enforced at decryption time and remains indistinguishable from standard training, guaranteeing private, autonomous, and verifiable FU in realistic FL deployments. To further enhance unlearning activity, we incorporate regularization objectives that suppress observable deviations in both the model's parameter space and its behavior on retained data — making unlearning \emph{indistinguishable} from normal training cycles.~\method~is agnostic to the underlying unlearning algorithm, serving as a simple yet effective mechanism to enable \emph{function hiding} and \emph{enforcement} in any FU scheme that relies on targeted updates. Our main contributions are as follows:

\begin{itemize}
    \item We introduce~\method, a FU approach that empowers clients to cryptographically enforce their ``\emph{right to be forgotten}'' — eliminating the need for server trust or cooperation.
    \vspace{2pt}
    \item We leverage functional encryption to make unlearning updates \emph{indistinguishable} from standard training updates — concealing their intent and preventing the server from identifying or altering them.
    \vspace{2pt}
    \item Our evaluation on $3$ datasets and $2$ architectures across diverse FL settings showcases~\method's~ability to achieve effective \emph{sample}-, \emph{class}-, and \emph{task}-level unlearning, with enforced deletion and preserved indistinguishability across model, behavioral, and communication patterns.
    \vspace{2pt}
    \item We demonstrate that~\method~serves as a \emph{drop-in} mechanism for any client-side FU method issuing targeted unlearning updates — enabling secure and enforceable unlearning without modifying existing training logic.
\end{itemize}

\section{Related Work}

\subsection{Machine Unlearning}

Machine Unlearning was first introduced in~\cite{cao2015mu}, predating legal frameworks like the ``\emph{right to be forgotten}'', as a means to efficiently remove the influence of specific data from trained models without full model retraining by transforming learning algorithms into summation-based forms. With respect to the modification scope, MU approaches can be broadly categorized into two main directions: \emph{data-based}, which restructure the training set to enable efficient forgetting, and \emph{model-based}, which directly modify the trained model to erase specific data influences. Data-based methods typically involve data pruning~\cite{sisa, chen2021} or synthetic data replacement~\cite{cao2015mu, cha2024learning, Shibata2021-mz}, followed by model retraining. In contrast, model-based techniques operate directly on the model itself, using selective fine-tuning~\cite{10.1007/s10994-022-06178-9}, parameter importance regularization — such as Elastic Weight Consolidation (EWC)~\cite{Devalapally2024-cg} and Memory Aware Synapses (MAS)~\cite{cha2024learning} — or adversarial perturbations in the representation space to suppress memorized information~\cite{Zhang2022-kx,9157084}. MU methods can also be categorized by the \emph{granularity} of forgetting: \emph{class-wise} unlearning targets entire semantic categories~\cite{Shibata2021-mz, 9157084}, whereas \emph{sample-wise} unlearning removes the influence of specific user-contributed samples~\cite{cha2024learning, cao2015mu, sisa}. Class-wise approaches often rely on selective retraining or class-specific masking, while sample-wise methods require more precise model adjustments and are often regarded as the general case due to their broader applicability~\cite{10.1145/3603620}. 

Beyond unlearning itself, recent works have explored \emph{verification} mechanisms to certify whether data has truly been forgotten. These include membership inference~\cite{10.1145/3460120.3484756}, influence-based auditing~\cite{xu2024reallyunlearnedverifyingmachine}, and black-box detectors that leverage auxiliary models to verify removal without access to the training pipeline~\cite{Wang2025-wu}. Nonetheless, all aforementioned approaches assume a centralized setting, where access to the full training dataset allows for direct control over both the unlearning and verification processes — assumptions that do not hold in federated settings. In FL, the distributed nature of data complicates both operations, and more critically, none of these centralized methods consider the need for \emph{enforcement} — i.e., ensuring faithful execution of unlearning requests without the possibility of omission or interference — which is a core requirement in trustless FL deployments.

\subsection{Federated Unlearning}

Federated Unlearning (FU) extends the concept of MU to decentralized settings, enabling clients to request the removal of their data contributions from a collaboratively trained model. Here, we can categorize FU schemes as either \emph{passive}, where the server approximates unlearning from historical updates without client involvement~\cite{liu2021federaser, fast}, or \emph{active}, where clients participate in a new training phase to support data removal~\cite{halimi2022federated, zhong2025unlearning, pan2025ga, 9964015}, often using tailored optimization strategies — such as Projected Gradient Descent (PGD)~\cite{halimi2022federated}, Elastic Weight Consolidation (EWC)~\cite{9964015} or Gradient Ascent (GA)~\cite{pan2025ga} — to selectively undo the influence of forgotten data. Beyond the mode of participation, FU also expands the traditional granularity of unlearning — typically class-level~\cite{10.1145/3485447.3512222, 10.1145/3657290} and sample-level~\cite{pan2025ga, zhong2025unlearning} — to include \emph{client-level} forgetting~\cite{9521274, halimi2022federated, 10735680}, reflecting the structured nature of federated participation. To ensure the fidelity of unlearning, VeriFi~\cite{verifi} introduces \emph{verification} protocols based on model discrepancy and posterior shifts to grant clients the ability to audit whether their data has been effectively removed from the global model. 

Nevertheless, existing FU methods offer no guarantee that unlearning requests will be executed and do not conceal the client’s intent to unlearn from the server. They rely on server-side cooperation — assuming the server will faithfully incorporate unlearning updates during aggregation — which may not hold in realistic FL deployments, leaving unlearning unenforced. We address this gap with~\method, a drop-in mechanism that enables \emph{enforceable} and \emph{functionally hidden} unlearning via encrypted client updates and secure aggregation.~\method~can be integrated into any FU approach based on client-side unlearning~\cite{halimi2022federated, liu2021federaser, 9521274, zhong2025unlearning, pan2025ga}, providing cryptographic guarantees that deletion requests are both private and honored on the server-side.

\section{Preliminaries}

\subsection{Efficient Decentralized Multi-Client FE (DMCFE)}

Decentralized Multi-Client Functional Encryption (DMCFE)~\cite{chotard2018decentralized} provides a cryptographic foundation for secure collaborative computation without relying on a trusted third-party authority (TPA). Its design is particularly well-suited to FL scenarios~\cite{9860595, 10064312, tsouvalas2025enccluster, tsouvalas2024enccluster} due to two key properties: (i) decentralized generation of partial functional keys, allowing clients to contribute to decryption without exposing their inputs, and (ii) ciphertext-key binding, which ensures decryption keys are usable only with their intended ciphertexts. In our work, we utilize the efficient DMCFE scheme introduced in~\cite{tsouvalas2024enccluster, tsouvalas2025enccluster}, which mitigates the computational and communication overhead of FE by applying weight clustering to compress model updates before encryption. This reduces input dimensionality and enables scalable secure aggregation with minimal client-side cost, making DMCFE practical even in resource-constrained FL environments while preserving strong cryptographic guarantees. 

We provide essential mathematical expressions of DMCFE's core functions. Let $\mathcal{F}$ be a family of functions $f$, and let $\mathcal{M}$ be a set of labels that bind each ciphertext to its intended functional key. A DMCFE scheme over $\mathcal{F}$ and labels $\mathcal{M}$ is a tuple of seven algorithms \resizebox{0.6\linewidth}{!}{$\mathcal{E}_{DMCFE} = (\mathsf{Setup}, \mathsf{KeyGen}, \mathsf{dKeyShare}, \mathsf{dKeyComb}, \mathsf{Cluster}, \mathsf{Enc}, \mathsf{Dec})$}:

\begin{itemize}
    \item $\mathsf{Setup}(\lambda, n)$: Takes as input a security parameter $\lambda$ and the number of clients $n$ and generates public parameters $\mathsf{pp}$. We assume that all the remaining algorithms implicitly contain $\mathsf{pp}$.
    \vspace{3pt}
    \item $\mathsf{KeyGen}(\mathsf{id}_i)$: Takes as input a client-specific identifier $\mathsf{id}_i$ and outputs a secret key $\mathsf{sk}_i$ and an encryption key $\mathsf{ek}_i$, unique to client $i$.
    \vspace{3pt}
    \item $\mathsf{dKeyShare}(\mathsf{sk}_i, f)$: Takes as input a secret key $\mathsf{sk}_i$ and a function $f \in \mathcal{F}$, and computes a partial functional decryption key $\mathsf{dk}_i$.
    \vspace{3pt}
    \item $\mathsf{dKeyComb}(\{\mathsf{dk}_i\}_{i \in \mathcal{N}})$: Takes as input a set of $n$ partial functional decryption keys $\{\mathsf{dk}_i\}_{i \in \mathcal{N}}$ and outputs the functional decryption key $\mathsf{dk}_f$.
    \vspace{3pt}
    \item $\mathsf{Cluster}(\theta, \kappa)$: Takes as input a model update vector $\theta \in \mathbb{R}^d$ and the number of clusters $\kappa$, and outputs a set of centroids $\mathcal{Z} = \{z_1, \dots, z_\kappa\}$ and a mapping matrix $\mathbf{P} \in \{1, \dots, \kappa\}^d$, where each entry $\mathbf{P}_i$ indicates that $\theta_i$ is assigned to centroid $z_{\mathbf{P}_i} \in \mathcal{Z}$.
    \vspace{3pt}
    \item $\mathsf{Enc}(\mathsf{ek}_i, x'_{i}, m)$: Takes as input the encryption key $\mathsf{ek}_i$, the compressed model update $x'_{i}$, and a label $m \in \mathcal{M}$, and outputs the ciphertext $\mathsf{ct}_{i,m}$.
    \vspace{3pt}
    \item $\mathsf{Dec}(\mathsf{dk}_f, \{\mathsf{ct}_{i,m}\}_{i \in \mathcal{N}})$: Takes as input a functional decryption key $\mathsf{dk}_f$ and ciphertexts under the same label $m$, and returns the result of computing $f$ over the encrypted inputs.
\end{itemize}

The $\mathsf{Cluster}$ step is implemented using K-means clustering, which minimizes the objective in Eq.~\eqref{eqn:wc} (see Appx.~\ref{asec:wc}).

\subsection{Measuring Weight Importance with MAS}\label{asec:mas}

To compute parameter importance scores \( \Omega \), Memory Aware Synap-ses (MAS)~\cite{mas} estimates how sensitive each parameter is to the model's output. Intuitively, parameters that induce large changes in the output when perturbed are considered more important. Formally, given a set of input samples \( \mathcal{D} = \{x_i\}_{i=1}^{|\mathcal{D}|} \), the importance of parameter \( \theta_j \) is computed as the average squared  \( \ell_2 \)-norm of the output gradient with respect to \( \theta_j \):

\begin{equation}\label{eqn:mas}
\Omega_j = \frac{1}{|\mathcal{D}|} \sum_{i=1}^{|\mathcal{D}|} \left\| \frac{\partial \| p_{›}(x_i) \|_2^2 }{\partial \theta_j} \right\|
\end{equation}

\noindent Here, \( \Omega_j \) reflects the contribution of weight \( \theta_j \) to the output activations across all inputs in \( \mathcal{D} \).

\section{Methodology}

\subsection{Problem Formulation}

We consider an FL setting with a server $S$ and a set of clients \( \mathcal{N} \). Each client \( i \in \mathcal{N} \) holds a private dataset \( \mathcal{D}_i \), which has previously contributed to training a model \( p_{\theta} \), parameterized by \( \theta \in \mathbb{R}^d \), in an FL manner. After participating in training, a client may retain a subset \( \mathcal{D}_i^l \subset \mathcal{D}_i \) and request unlearning of a disjoint subset \( \mathcal{D}_i^u \), aiming to erase its influence from the model. Formally, we have the following objective function:

\begin{equation}\label{eqn:pf}
    \resizebox{0.65\linewidth}{!}{
        $
        \begin{aligned}
                \theta^* = \arg\min_{\theta} \sum_{i \in \mathcal{N}}  \left( \sum_{(x, y) \in \mathcal{D}_i^l} \mathcal{L}_{\text{CE}}(p_{\theta}(x), y) + \sum_{(x, y) \in \mathcal{D}_i^u} \mathcal{L}_{\text{UN}}(p_{\theta}(x), y) \right)
        \end{aligned}$
    }
\end{equation}

\noindent  where \( \mathcal{L}_{\text{CE}} \) denotes the cross-entropy loss, and \( \mathcal{L}_{\text{UN}} \) the unlearning loss applied to samples in \( \mathcal{D}_i^u \).

\subsection{Threat Model and Objectives.}

We assume an \textit{Honest-but-Curious} (\textit{HbC}) server that adheres to the FL protocol, but may additionally attempt to infer the identity of clients submitting updates intended for unlearning (i.e., requesting the removal of their data contributed to the model), and/or disregard the unlearning request by excluding their updates from aggregation. Clients are trusted to perform key generation, encryption, and unlearning operations correctly, while no trusted TPA is assumed to assist with or mediate in the process. 

Our \emph{objectives} are to: (i) enable clients to initiate deletion of their prior contributions, (ii) ensure the global model incorporates this deletion, and (iii) prevent the server from identifying the requesting client, detecting that an unlearning request occurred, or inferring its content.

\subsection{\method: Enforceable Federated Unlearning}

\noindent \textbf{\texttt{EFU}~Overview.} 
Clients begin by generating key pairs \( (\mathsf{sk}_i, \mathsf{ek}_i) \) using \( \mathsf{KeyGen}(\mathsf{id}_i) \). At the start of each training round \( r \), each client computes a partial decryption key \( \mathsf{dk}_i^r \) via \( \mathsf{dKeyShare} \). The client then performs local training — either to \emph{learn} from or \emph{unlearn} specific data — and compresses the resulting model update using \( \mathsf{Cluster} \) to produce a centroid set \( \mathcal{Z} \) and a mapping matrix \( \mathbf{P} \in \{1, \dots, \kappa\}^d \). The centroids \( \mathcal{Z} \) are encrypted under round label \( r \) using \( \mathsf{Enc} \), yielding \( \mathcal{Z}_{\text{enc}} \), and the triplet \( (\mathcal{Z}_{\text{enc}}, \mathbf{P}, \mathsf{dk}_i^r) \) is sent to the server. On the server-side, each client's encrypted update \( \theta_{\text{enc}} \) is constructed by expanding the encrypted centroids \( \mathcal{Z}_{\text{enc}} \) using the mapping matrix \( \mathbf{P} \). The server then uses $\mathsf{dKeyComb}$ to combine the partial decryption keys into $\mathsf{dk}_f$, and utilize $\mathsf{Dec}$ to securely aggregate the encrypted updates. Fig.~\ref{fig:overview} provides an overview of~\method's training process, while Algorithm~\ref{alg:method} outlines the~\method~algorithm.

\subsubsection{\textbf{Client-Side Operation.}} We now focus on the construction of model updates on the client-side. Here, each client constructs its local update either by performing \textit{learning on retained data} or by executing \textit{unlearning on designated forget data} (i.e., remove its influence from the model).

\noindent \textbf{\textit{Local Learning}}: Given a retained dataset \( \mathcal{D}_{i}^{l} \), the client performs standard local training using the cross-entropy loss \( \mathcal{L}_{\text{CE}} \). Specifically, the model is updated via multiple steps of mini-batch stochastic gradient descent (SGD), yielding a model update \( \theta_i^r \), which is then compressed and encrypted prior to transmission.

\begin{algorithm}[!ht]
    \centering \footnotesize
    \caption{\small{\method: Enforced Federated Unlearning via Functional Encryption. $\eta$ is the learning rate, while $\kappa$ refer to the number of clusters during model updates compression. \textit{SecureAggr} refers to the secure aggregation procedure in Algorithm~\ref{alg:secure_aggr}.}}\label{alg:method}
    \begin{algorithmic}[1]
        {\setlength{\baselineskip}{11pt}}
        \State \textit{ClientInit}($\lambda$, $\mathcal{N}$)
        \State Server $\mathcal{S}$ initializes model parameters $\theta$ and computes total data size $|D|{=}\sum_{i \in \mathcal{N}} |\mathcal{D}_i|$
        \For{$r=1$ to $R$}
            \For{$i \in \mathcal{N}$ \textbf{in parallel}}
                \State $\mathsf{dk}_i^r \gets \mathsf{dKeyShare}(\mathsf{sk}_i, f)$
                \State $\theta_i^r \gets$ \textit{ClientUpdate}($\theta^r$, $\mathcal{D}^l_i$, $\mathcal{D}^u_i$)
                \State $(\mathcal{Z}^r_i, \mathbf{P}^r_i) \gets \mathsf{Cluster}(\theta_i^r, \kappa)$
                \State $\mathcal{Z}^{r}_{i,\text{enc}} \gets \left\{ \mathsf{Enc}(\mathsf{ek}_i, \mathcal{Z}^r_i[j], r) \right\}_{j=1}^{\kappa}$
            \EndFor
            \State $\theta^{r+1} \gets \textit{SecureAggr} \left( \left\{ (\hat{\mathcal{Z}}^r_i, \mathbf{P}^r_i, \mathsf{dk}_i^r \right\}_{i \in \mathcal{N}} \right)$
        \EndFor
        \vspace{5pt}
        \Procedure{\textit{ClientInit}}{$\lambda,~\mathcal{N}$}
            \State $\mathsf{pp} \gets \mathsf{Setup}(\lambda, |\mathcal{N}|)$
            \For{$i \in \mathcal{N}$ \textbf{in parallel}}
                \State $(\mathsf{ek}_i, \mathsf{sk}_i) \gets \mathsf{KeyGen}(\mathsf{id}_i)$
            \EndFor
        \EndProcedure
        \vspace{5pt}
        \Procedure{\textit{ClientUpdate}}{$\theta,~\mathcal{D}^l,~\mathcal{D}^u$}
            \If{$\mathcal{D}^u = \emptyset$} \Comment{\textbf{\textit{Local learning}}}
                \For{batch $b \in \mathcal{D}^l$}
                    \State $\theta \gets \theta - \eta \cdot \nabla_\theta \mathcal{L}_{\text{CE}}(b)$
                \EndFor
            \Else \Comment{\textbf{\textit{Local unlearning}}}
                \State $\mathcal{D}^{\text{u,adv}} \gets \text{$\ell_2$-PGD}(\mathcal{D}^u)$ \Comment{\textit{Run once, offline before round begins}}
                \vspace{2pt}
                \For{batch ($b$, $b^{\text{adv}}$) $\in (\mathcal{D}^u,~\mathcal{D}^{\text{u,adv}})$}
                    \State \resizebox{0.5\linewidth}{!}{$\theta^{*} \gets \theta - \eta \cdot \nabla_\theta \left( -\mathcal{L}_{\text{CE}}(b) + \mathcal{L}_{\text{CE}}(b^{\text{adv}}) + \sum_{j=1}^{d} \bar{\Omega}_j (\theta_j - \tilde{\theta}_j)^2 \right)$} \Comment{\textit{Eq.~\eqref{eq:unlearn}}}
                \EndFor
            \EndIf
            \State \Return $\theta^{*}$
        \EndProcedure
    \end{algorithmic}
\end{algorithm}

\noindent \textbf{\textit{Local Unlearning}}: When a client opts to remove the influence of \( \mathcal{D}_i^u \) from the model \( p_{\theta} \), we adopt \emph{sample-wise} unlearning based on gradient ascent~\cite{cha2024learning, halimi2022federated, pan2025ga}. While we use this approach for its effectiveness, it is important to note that~\method~serves as a \emph{drop-in mechanism} for any FU scheme that issues client-side unlearning updates~\cite{halimi2022federated, liu2021federaser, 9521274, zhong2025unlearning, pan2025ga}, enabling \emph{enforceable} and \emph{indistinguishable} FU. The key idea behind gradient-ascent unlearning is to reverse the learning signal by applying a negated cross-entropy loss on the forget set, \( -\mathcal{L}_{\text{CE}}(\mathcal{D}_i^u) \), thereby encouraging the model to misclassify these samples. 

Although this objective suppresses memorized information, it may introduce detectable shifts in model behavior on the retained data \( \mathcal{D}_i^l \), potentially revealing that unlearning has occurred~\cite{cha2024learning, pan2025ga}. To mitigate this, we incorporate the auxiliary adversarial training mechanism from~\cite{cha2024learning}, which stabilizes the model’s internal representations by applying a standard \( \mathcal{L}_{\text{CE}} \) loss on adversarial variants \( \mathcal{D}_i^{\text{u,adv}} \) of the forget set. We generate these examples \emph{once} on the client-side, before the unlearning round begins (i.e., offline), using targeted \( \ell_2 \)-PGD attacks~\cite{pgd} with randomly chosen incorrect labels. Furthermore, to minimize \emph{observable} parameter drift during unlearning, we apply a MAS-based regularization that estimates parameter importance from \( \mathcal{D}_i^u \) and penalizes updates to parameters less relevant for ``\textit{forgetting}'' \( \mathcal{D}_i^u \) (see Sec.~\ref{asec:mas}). Intuitively, this regularization makes the resulting update pattern harder to distinguish from normal learning, thereby concealing unlearning traces in model's parameter space.

Formally, the optimization function for local unlearning of \( \mathcal{D}_i^u \) is:

\begin{equation}\label{eq:unlearn}    
    \resizebox{0.6\linewidth}{!}{
        $
        \begin{aligned}
    \mathcal{L}_{\text{UN}} = \underbrace{-\mathcal{L}_{\text{CE}}(\mathcal{D}_i^u)}_{\textit{forget loss} \vphantom{\textstyle \sum_{j=1}^{d} (\theta_j - \tilde{\theta}_j)^2}} + \underbrace{\mathcal{L}_{\text{CE}}(\mathcal{D}_i^{\text{u,adv}})}_{\textit{feature collapse aux. loss} \vphantom{\textstyle \sum_{j=1}^{d} (\theta_j - \tilde{\theta}_j)^2}} + \underbrace{\textstyle \sum_{j=1}^{d} \bar{\Omega}_j (\theta_j - \tilde{\theta}_j)^2}_{\textit{parameter drift reg.} \vphantom{\big|}}
        \end{aligned}$
    }
\end{equation}

\noindent where \( \mathcal{D}_i^u \) is the client’s forget set, and \( \mathcal{D}_i^{\text{u,adv}} \) consists of adversarial examples generated from \( \mathcal{D}_i^u \) using targeted \( \ell_2 \)-PGD attacks with randomly selected incorrect labels. \( \tilde{\theta}_j \) denotes the model's parameters \( \theta_j \) prior to unlearning, while \( \bar{\Omega}_j = 1 - \Omega_j \) is the inverted MAS score that prioritizes updates to parameters most relevant for ``\textit{forgetting}'' \( \mathcal{D}_i^u \). Hence, during unlearning, each client optimizes \( \mathcal{L}_{\text{UN}} \), as depicted in Eq.~\eqref{eq:unlearn}, to generate a model update \( \theta_i^r \) that forgets its designated data \( \mathcal{D}_i^u \), which is then compressed and encrypted before upload to the server.

\noindent \textbf{\textit{Binding Updates to Aggregation via Encryption}}: Upon completing either a learning or unlearning update, each client compresses its local model weights \( \theta_i^r \in \mathbb{R}^d \) into a set of \( \kappa \) representative centroids \( \mathcal{Z}_i^r \) ($|\mathcal{Z}_i^r|$=$\kappa$) and a mapping matrix \( \mathbf{P}_i^r \in \{1, \dots, \kappa\}^d \) that assigns each parameter to a designated centroid. We then encrypt \( \mathcal{Z}_i^r \) into \( \mathcal{Z}_{i,\text{enc}}^r \) using \( \mathsf{ek}_i \) under the round label \( r \), and send it to server-side. This compression via weight clustering preserves the semantic direction of updates while drastically reducing encryption overhead~\cite{tsouvalas2025enccluster}; intuitively, it shifts costly cryptographic operations from full model dimensionality to just $\kappa$ centroids (where $d \gg \kappa$). More importantly, by encrypting client updates and binding them to round-specific aggregation operations, the server is prevented from altering, skipping, or distinguishing unlearning from learning—thereby enforcing unlearning while concealing its occurrence. We formally analyze these properties in Sec.~\ref{sec:privacy-main}.

\subsection{Server-side Secure Aggregation}\label{ssec:server}

After each training round \( r \), the server \( \mathcal{S} \) aggregates the encrypted model updates as follows: it first reconstructs the encrypted model update \( \theta_{i,\text{enc}}^{r} \) by expanding centroids \( \mathcal{Z}_{i,\text{enc}}^r \) using the mappings \( \mathbf{P}_i^r \). It then derives the aggregate functional key \( \mathsf{dk}_f \) via \( \mathsf{dKeyComb}(\{ \mathsf{dk}_i^r \}) \). Since \( \mathsf{dk}_f \) is cryptographically bound to both the aggregation logic (i.e., function $f$) and the encrypted inputs, the server is forced to execute the computation exactly as intended — any deviation would render the decryption invalid (see Sec. \ref{sec:privacy-main} and Theorem~\ref{thm:fhe} in Appx.~\ref{app:proofs}). Subsequently, secure aggregation proceeds as:

\begin{equation}\label{eq:secure_agg}
    \theta^{r+1} = \left \{ \mathsf{Dec} \left( \{ \theta_{i,\text{enc}}^{r}[j] \}_{i \in \mathcal{N}}, \mathsf{dk}_f \right ) \right \}_{j=1}^{d},
\end{equation}

\noindent where \( \mathsf{Dec} \) returns the decrypted aggregated weight for each parameter index \( j \); yielding the updated global model \( \theta^{r+1} \), which is then shared with clients for the next training round. We apply weighted aggregation as in FedAvg~\cite{konevcny2016federated}, where each client scales its encrypted centroids by its local-to-total sample ratio prior to encryption, following~\cite{tsouvalas2025enccluster}. The detailed algorithm of secure aggregation is presented in Algorithm \ref{alg:secure_aggr}.

\begin{algorithm}[!t]
    \centering \footnotesize
    \caption{\small{\textit{SecureAggr}: Server-side secure aggregation over encrypted client updates.}}\label{alg:secure_aggr}
    \begin{algorithmic}[1]
        \State \textbf{Inputs:} Encrypted centroids $\left\{ \mathcal{Z}_{i,\text{enc}}^r \right\}_{i \in \mathcal{N}}$, mapping matrices $\left\{ \mathbf{P}_i^r \right\}_{i \in \mathcal{N}}$, partial decryption keys $\{\mathsf{dk}_i^r\}_{i \in \mathcal{N}}$
        \vspace{2pt}
        \State \textbf{Output:} Aggregated model $\theta^{r+1}$
        \vspace{2pt}
        \State $\mathsf{dk}_f \gets \mathsf{dKeyComb}(\{\mathsf{dk}_i^r\}_{i \in \mathcal{N}})$
        \vspace{2pt}
        \For{$i \in \mathcal{N}$}
            \State $\theta_{i,\text{enc}}^r \gets \textsc{Expand}(\mathcal{Z}_{i,\text{enc}}^r, \mathbf{P}_i^r)$ \Comment{\textit{// Expand enc. centroids to full model size}}
        \EndFor
        \vspace{2pt}
        \State $\theta^{r+1} \gets \left\{ \mathsf{Dec} \left( \left\{ \theta_{i,\text{enc}}^r[j] \right\}_{i \in \mathcal{N}},~\mathsf{dk}_f \right) \right\}_{j=1}^{d}$ \Comment{\textit{// Eq. \eqref{eq:secure_agg}}}
        \vspace{2pt}
        \State \Return $\theta^{r+1}$
    \end{algorithmic}
\end{algorithm}

\subsection{Unlearning Guarantees} \label{sec:privacy-main}

We now analyze how~\method~satisfies two core guarantees in the context of FU: (i) \textit{enforceable unlearning}, and (ii) \textit{indistinguishability of update type} — both ensured under the IND-security of DMCFE against chosen-plaintext attacks~\cite{chotard2018decentralized}. We refer readers to Appx.\ref{app:proofs} for formal definitions and proofs.

\noindent \textbf{\textit{Enforceable unlearning}}: Each client encrypts its update under a round-specific label \( r \), binding it to a fixed aggregation function \( f \). The decryption key \( \mathsf{dk}_f \), constructed via $\mathsf{dKeyComb}$ from label-bound partial keys, ensures that only the complete, authorized aggregation over all ciphertexts for label \( r \) can be decrypted (Eq.\eqref{eq:secure_agg}). Any omission, modification, or partial aggregation would invalidate decryption. This guarantee is ensured by the IND-security of the DMCFE scheme, which enforces that only complete, label-consistent aggregations can be successfully decrypted (see Theorem~\ref{thm:fhe} in Appx.~\ref{app:proofs}).

\noindent \textbf{\textit{Update-type indistinguishability}}: Learning and unlearning updates follow the same computational pipeline--local training, clustering, encryption--under the same label and key structure, producing ciphertexts of identical size and format. The IND-security of DMCFE guarantees that no adversary with encryption and partial key access (excluding full collusion) can distinguish between ciphertexts corresponding to semantically different updates (e.g., learn vs.\ unlearn), as long as the underlying function outputs remain indistinguishable (see Theorem~\ref{thm:uti} in Appx.~\ref{app:proofs}). We further provide a quantitative analysis across multiple model-based and system-based detectability indicators in Sec.~\ref{ssec:unlearn_hide}.
\vspace{6pt}

\noindent While~\method~ensures unlearning is cryptographically enforced, assessing its behavioral effect on the model's predictions after forgetting is also crucial; yet the non-convex and stochastic nature of FL precludes exact guarantees on data removal influence \cite{pan2025ga}. Empirically, we estimate residual influence by comparing the model \( p_{\theta} \) after unlearning to a counterfactual model \( p_{\theta_{\setminus \mathcal{D}^u}} \) retrained without access to \( \mathcal{D}^u \), using the average prediction gap \( \varepsilon \) over held-out data as a practical proxy. We refer reader to Appx.~\ref{asec:mit} for formal definitions of forgetting metrics in MU. 

\begin{table*}[!t]
    \centering 
    \caption{Unlearning performance evaluation under \textbf{\textit{sample-wise}} and \textbf{\textit{class-wise}} forgetting across CIFAR-10, LFW, and FEMNIST using (a) \emph{ResNet-18}~\cite{resnet} and (b) \emph{ViT-B/32}~\cite{vit}. We report accuracy on retained ($\text{Acc}$) and forgotten data ($\text{Acc}_f$) on the test set. FL parameters: $N{=}10$, $N_u{=}1$, $R{=}100$, $R_u{=}50$, and $\rho{=}0.2$.}
    \label{tab:main_res}
    \begin{minipage}[t]{0.49\textwidth}
        \centering \small
        \resizebox{\linewidth}{!}{
            \begin{tabular}{@{}clcccccc@{}}
                \toprule
                \multicolumn{1}{c}{\multirow{3}{*}{\shortstack{\textbf{Forgetting}\\\textbf{Scenario}}}} 
                & \multicolumn{1}{c}{\multirow{2}{*}{\textbf{Method}}} 
                & \multicolumn{2}{c}{\textbf{CIFAR-10}} 
                & \multicolumn{2}{c}{\textbf{LFW}} 
                & \multicolumn{2}{c}{\textbf{FEMNIST}} \\
                \cmidrule(lr){3-4} \cmidrule(lr){5-6} \cmidrule(lr){7-8}
                & & $\text{Acc}$ & $\text{Acc}_f$ (\textcolor{green}{$\downarrow$}) & $\text{Acc}$ & $\text{Acc}_f$ (\textcolor{green}{$\downarrow$}) & $\text{Acc}$ & $\text{Acc}_f$ (\textcolor{green}{$\downarrow$}) \\ \midrule
                \multirow{9}{*}{\shortstack{\textit{Sample-wise} \\ ($\gamma_s{=}0.1$)}}
                    & \textbf{Full Retrain (FedAvg)}                & 93.76 & --    & 84.53 & --    & 77.23 & --   \\ 
                    \cmidrule(lr){2-8}
                    & \textbf{SGA-EWC~\cite{9964015}}               & 91.32 & --    & 80.27 & --    & 75.12 & --   \\ 
                    & \textbf{PGD~\cite{halimi2022federated}}       & 91.84 & --    & 80.92 & --    & 75.37 & --   \\
                    & \textbf{FedOSD~\cite{pan2025ga}}              & 92.97 & --    & 83.78 & --    & 76.44 & --   \\
                    \cmidrule(lr){2-8}
                    & \textbf{\method$_\mathrm{SGA-EWC}$}           & 91.01 & --    & 79.97 & --    & 74.82 & -- \\
                    & \textbf{\method$_\mathrm{PGD}$ }              & 91.56 & --    & 80.04 & --    & 74.97 & -- \\
                    & \textbf{\method$_\mathrm{FedOSD}$}            & \textbf{92.61} & --    & \textbf{83.26} & --    & 76.01 & -- \\ 
                    & \textbf{\method}                              & 92.43 & --    & 83.13 & --    & \textbf{76.33} & -- \\
                    \cmidrule(lr){1-8}
                \multirow{9}{*}{\shortstack{\textit{Class-wise} \\ ($\gamma_c{=}0.1$)}}
                    & \textbf{Full Retrain (FedAvg)}                & 94.82 & 13.72 & 85.11 & 3.78  & 77.69 & 1.54 \\
                    \cmidrule(lr){2-8}
                    & \textbf{SGA-EWC~\cite{9964015}}               & 91.23 & 14.69 & 80.03 & 8.55  & 73.80 & 4.85 \\
                    & \textbf{PGD~\cite{halimi2022federated}}       & 91.82 & 14.31 & 80.55 & 7.26  & 74.52 & 4.23 \\
                    & \textbf{FedOSD~\cite{pan2025ga}}              & 93.06 & 12.77 & 83.77 & 6.01  & 75.40 & 2.49 \\
                    \cmidrule(lr){2-8}
                    & \textbf{\method$_\mathrm{SGA-EWC}$}           & 90.91 & 14.71 & 79.72 & 8.52  & 73.55 & 4.82 \\
                    & \textbf{\method$_\mathrm{PGD}$}               & 91.61 & 14.29 & 80.24 & 7.23  & 74.20 & 4.17 \\
                    & \textbf{\method$_\mathrm{FedOSD}$}            & \textbf{92.73} & \textbf{12.54} & 83.41 & \textbf{6.05}  & 75.02 & 2.39 \\
                    & \textbf{\method}                              & 92.43 & 13.06 & \textbf{83.78} & 6.47  & \textbf{75.93} & \textbf{1.97} \\
                \bottomrule
            \end{tabular}
        }
        \par\vspace{6pt}\textbf{(a) ResNet-18}
    \end{minipage}%
    \hfill
    \begin{minipage}[t]{0.49\textwidth}
        \centering \small
        \resizebox{\linewidth}{!}{
            \begin{tabular}{@{}clcccccc@{}}
                \toprule
                \multicolumn{1}{c}{\multirow{3}{*}{\shortstack{\textbf{Forgetting}\\\textbf{Scenario}}}} 
                & \multicolumn{1}{c}{\multirow{2}{*}{\textbf{Method}}} 
                & \multicolumn{2}{c}{\textbf{CIFAR-10}} 
                & \multicolumn{2}{c}{\textbf{LFW}} 
                & \multicolumn{2}{c}{\textbf{FEMNIST}} \\
                \cmidrule(lr){3-4} \cmidrule(lr){5-6} \cmidrule(lr){7-8}
                & & $\text{Acc}$ & $\text{Acc}_f$ (\textcolor{green}{$\downarrow$}) & $\text{Acc}$ & $\text{Acc}_f$ (\textcolor{green}{$\downarrow$}) & $\text{Acc}$ & $\text{Acc}_f$ (\textcolor{green}{$\downarrow$}) \\ \midrule
                \multirow{9}{*}{\shortstack{\textit{Sample-wise} \\ ($\gamma_s{=}0.1$)}}
                    & \textbf{Full Retrain (FedAvg)}             & 98.21 & --    & 90.54 & --    & 84.11 & --   \\ 
                    \cmidrule(lr){2-8}
                    & \textbf{SGA-EWC~\cite{9964015}}            & 94.50 & --    & 87.02 & --    & 80.49 & --   \\ 
                    & \textbf{PGD~\cite{halimi2022federated}}    & 96.29 & --    & 88.71 & --    & 82.09 & --   \\
                    & \textbf{FedOSD~\cite{pan2025ga}}           & 97.31 & --    & 90.10 & --    & 83.04 & --   \\
                    \cmidrule(lr){2-8}
                    & \textbf{\method$_\mathrm{SGA-EWC}$}        & 94.10 & --    & 86.52 & --    & 80.02 & -- \\
                    & \textbf{\method$_\mathrm{PGD}$}            & 96.07 & --    & 88.79 & --    & 81.94 & -- \\
                    & \textbf{\method$_\mathrm{FedOSD}$}         & \textbf{96.89} & --    & \textbf{89.57} & --    & 82.64 & -- \\ 
                    & \textbf{\method}                           & 96.55 & --    & 89.49 & --    & \textbf{82.79} & -- \\
                    \cmidrule(lr){1-8}
                \multirow{9}{*}{\shortstack{\textit{Class-wise} \\ ($\gamma_c{=}0.1$)}}
                    & \textbf{Full Retrain (FedAvg)}             & 98.91 & 12.89 & 93.34 & 4.71 & 84.76 & 1.67 \\
                    \cmidrule(lr){2-8}
                    & \textbf{SGA-EWC~\cite{9964015}}            & 94.51 & 15.45 & 87.02 & 8.25  & 80.49 & 5.15 \\
                    & \textbf{PGD~\cite{halimi2022federated}}    & 94.94 & 12.81 & 90.01 & 7.93  & 79.83 & 4.97 \\
                    & \textbf{FedOSD~\cite{pan2025ga}}           & 97.31 & 13.05 & 90.97 & 5.32  & 83.11 & 2.48 \\
                    \cmidrule(lr){2-8}
                    & \textbf{\method$_\mathrm{SGA-EWC}$}        & 94.10 & 15.75 & 86.52 & 8.21  & 80.02 & 5.17 \\
                    & \textbf{\method$_\mathrm{PGD}$}            & 94.70 & 13.44 & 89.77 & 7.91  & 79.50 & 4.94 \\
                    & \textbf{\method$_\mathrm{FedOSD}$}         & 96.89 & \textbf{13.17} & \textbf{90.51} & \textbf{5.44}  & \textbf{82.73} & 2.24 \\
                    & \textbf{\method}                           & \textbf{97.45} & 13.29 & 90.43 & 5.71  & 82.14 & \textbf{2.09} \\
                \bottomrule
            \end{tabular}
        }
        \par\vspace{6pt}\textbf{(b) ViT-B/32}
    \end{minipage}
\end{table*}

\section{Evaluation}\label{sec:eval}

\subsection{Experimental Results}

\noindent \textbf{Datasets \& Models.} We evaluate~\method~on publicly available vision classification datasets, namely CIFAR-10\cite{cifar} (10 object classes), FEMNIST~\cite{femnist} (62 handwritten character classes), and LFW~\cite{lfw} (30 identity classes). To further evaluate \emph{task-level} unlearning in a multi-task FL setting, we use LFW, where each client jointly performs identity and gender classification. For all datasets, we use ResNet-18~\cite{resnet} and ViT-B/32~\cite{vit} as backbone models, initialized with ImageNet pretrained weights from Hugging Face. All models are trained with a batch size of $64$ using the Adam optimizer with a learning rate of $10^{-3}$ and weight decay of $10^{-2}$. Model updates are compressed into $\kappa = 64$ clusters using weight clustering to minimize encryption overhead, following~\cite{tsouvalas2025enccluster}.
\vspace{3pt}

\noindent \textbf{Experimental Setup.} Our FL simulations are implemented using Flower~\cite{beutel2020flower}, with key parameters including: the total number of clients ($N$), total training rounds ($R$), local training epochs per round ($E$), client participation rate per round ($\rho$), and the number of weight clusters ($\kappa$). In addition to standard FL settings, we define unlearning-specific parameters: the number of unlearning clients ($N_u$), the round at which unlearning begins ($R_{\text{u}}$), the fraction of each client's samples designated for removal in \emph{sample-wise} forgetting ($\gamma_s$), and the fraction of classes to be forgotten in \emph{class-wise} scenarios ($\gamma_c$). In all unlearning scenarios, models are first trained for 50 rounds (i.e., until convergence), followed by 10 rounds of unlearning, after which clients either resume training on retained data or drop out if none remains. We set $E{=}1$ for standard learning and $E{=}5$ during the unlearning phase, using a 256-bit encryption key for DMCFE as in~\cite{tsouvalas2025enccluster}. In experiments with $\rho < 1.0$, clients are randomly selected in each round.

Data is split across clients using a Dirichlet distribution $Dir(a)$ \cite{li2022federated}, where $a$ controls class distribution. We consider non-IID settings, using $a=0.1$, leading to skewed label distributions --- typical in FL scenarios\cite{kairouz2021advances}. We fix the seed in data partitioning to ensure consistent data splits across experiments for direct comparison. All experiments run on NVIDIA RTX4090 GPUs in an internal cluster server with 96 CPU cores and one GPU per run.
\vspace{3pt}

\noindent \textbf{Baselines.} We evaluate two unlearning scenarios: (i) \emph{sample-wise}, where a fraction $\gamma_s$ of each client’s samples is randomly removed, and (ii) \emph{class-wise}, where a fraction $\gamma_c$ of classes is forgotten. We intentionally omit \emph{client-level} forgetting, as it represents a special case of the more general \emph{sample-wise} and \emph{class-wise} settings considered here. Instead, we further explore \emph{task-level} unlearning in multi-task FL scenarios, where one of several jointly learned tasks — via a shared backbone and distinct task-specific heads — is selectively removed. This setting aligns with emerging real-world deployments in which federated models are increasingly designed to support multiple tasks concurrently~\cite{tsouvalas2025matu}. To evaluate unlearning performance, we use test accuracy ($\text{Acc}$) to assess model performance on retained data, while in \textit{class-wise} settings, we further distinguish between accuracy on retained classes (denoted as $\text{Acc}$ for consistency) and on forgotten classes ($\text{Acc}_f$). 

Importantly, our goal is \textbf{\emph{not to achieve state-of-the-art unlearning performance}}, but to demonstrate that unlearning can---and should---be \textbf{\emph{cryptographically enforced and made indistinguishable}} during FL training via~\method. For this, in addition to our proposed unlearning loss (Eq.~\ref{eq:unlearn}), we also apply~\method~to several alternative FU optimization schemes that generate tailored client updates for forgetting, including PGD~\cite{halimi2022federated}, FedOSD~\cite{pan2025ga}, and SGA-EWC~\cite{9964015}, using their default hyperparameters. Here, we retain the cryptographic protocol of~\method~while adopting the unlearning optimization strategy of each respective method, incorporating the MAS regularization term into all approaches except SGA-EWC, which already utilizes EWC as a regularizer. We denote these integrated variants as \method$_\mathrm{PGD}$, \method$_\mathrm{FedOSD}$, and \method$_\mathrm{SGA\text{-}EWC}$, respectively. To ensure fair comparison, all methods perform unlearning over the same fixed number of rounds, and reported results are averaged over three independent runs.

\subsection{\method~Unlearning Performance}

\noindent \textbf{Sample-wise \& Class-wise Unlearning.} We start by evaluating~\method~on CIFAR-10, LFW, and FEMNIST under both \emph{sample-wise} and \emph{class-wise} forgetting across diverse model architectures and unlearning strategies. Our findings, reported in Table~\ref{tab:main_res}, highlight the efficacy of~\method~in achieving reliable forgetting across all scenarios. Specifically, our proposed unlearning loss yields strong performance, often matching or exceeding that of existing FU methods (e.g., on FEMNIST), while preserving accuracy on retained data comparable to full retraining — consistently across both ResNet-18 and ViT-B/32 models. This demonstrates~\method's capability to effectively remove client data influence in a wide range of FU settings. 

Furthermore, in the \method$_\mathrm{PGD}$, \method$_\mathrm{FedOSD}$, and \method$_\mathrm{SGA\text{-}EWC}$ rows of Table~\ref{tab:main_res}, where~\method~is integrated with alternative unlearning objectives, we observe minor performance degradation — due to the lossy compression from weight clustering — compared to their original counterparts. Notably, this gap can be further mitigated by increasing the number of clusters (i.e., $\kappa{>}64$), as demonstrated in~\cite{tsouvalas2025enccluster}, since secure aggregation yields exact results upon decryption. These findings confirm that~\method~can be seamlessly introduced as a \emph{drop-in mechanism} to cryptographically enforce unlearning without modifying the optimization strategies of existing FU schemes; thereby highlighting the broad applicability of~\method~across existing FU approaches.

\begin{wraptable}{r}{0.5\textwidth}
    \vspace{-13pt}
    \centering \small
    \caption{\emph{Task-level unlearning} in multi-task federated settings. We report test accuracy on the retained task (\textit{gender}, $\text{Acc}_\textit{gen}$) and the forgotten task (\textit{identity}, $\text{Acc}_\textit{id}$) on LFW using a ViT-B/32 as shared backbone with task-specific heads. Federated parameters are set to $N{=}30$, $N_u{=}30$, $R{=}100$, $R_u{=}50$, and $\rho{=}1.0$.}    
    \label{tab:multitask}
    \resizebox{0.9\linewidth}{!}{
        \begin{tabular}{@{}lcc@{}}
            \toprule
            \multicolumn{1}{c}{\textbf{Method}} & $\text{Acc}_\textit{gen}$ (\textbf{Retain}) & $\text{Acc}_\textit{id}$ (\textbf{Forget} / \textcolor{green}{$\downarrow$}) \\
            \midrule
            \textbf{Full Retrain (\textit{FedAvg})}      & 96.86 & 4.71 \\ 
            \midrule
            \textbf{SGA-EWC~\cite{9964015}}              & 93.54 & 19.47 \\
            \textbf{PGD~\cite{halimi2022federated}}      & 94.81 & 10.32 \\
            \textbf{FedOSD~\cite{pan2025ga}}             & 95.02 & 5.01 \\ 
            \midrule
            \textbf{\method$_\mathrm{SGA-EWC}$}          & 93.66 & 19.33 \\ 
            \textbf{\method$_\mathrm{PGD}$}              & 94.93 & 10.02 \\
            \textbf{\method$_\mathrm{FedOSD}$}           & 95.17 & \textbf{4.92} \\
            \textbf{\method}                             & \textbf{95.37} & 5.12 \\
            \bottomrule
        \end{tabular}%
    }
\end{wraptable}

\vspace{1pt}
\noindent \textbf{Task-level Unlearning.} We now explore a more structured unlearning scenario: \emph{task-level} unlearning in multi-task FL. This setting reflects real-world deployments where a single model supports multiple tasks simultaneously — for example, learning shared representations across classification objectives. To simulate this, we use LFW with ViT-B/32 as a shared backbone and distinct heads for face identification and gender classification. Initially, all clients collaboratively train the model on both tasks. At round $R_u{=}50$, a system-wide unlearning request is issued to remove the identity classification task, prompting all clients to unlearn face identification while retaining gender classification. To do so, we simultaneously optimize the standard cross-entropy loss for gender and~\method's unlearning loss (see Eq.~\ref{eq:unlearn}) for identity removal on each batch of client data.

As shown in Table~\ref{tab:multitask}, we observe that while all methods retain high accuracy on the preserved gender task (within 1.5\% of full retraining), their ability to forget the identity task varies significantly — especially when compared to sample- and class-wise unlearning (Table~\ref{tab:main_res}). Specifically, \method~and FedOSD achieve near-complete forgetting on the identity task, whereas PGD retains substantially more residual knowledge ($\approx$10\%) and SGA-EWC fails to remove task-specific information entirely ($\approx$20\%). Notably, all variants of~\method~that incorporate MAS regularization (i.e., \method, \method$_\mathrm{PGD}$, \method$_\mathrm{FedOSD}$) consistently outperform their base counterparts on both retention and forgetting, suggesting that MAS helps disentangle task-specific parameters by penalizing updates to weights irrelevant to the forgetting objective. More importantly, these results highlight that \emph{task-level unlearning} introduces novel challenges in multi-task FL due to shared backbone representations, with~\method~remaining an effective drop-in solution to \emph{enforce} unlearning without compromising utility.

\subsection{\method~Unlearning Indistinguishability}\label{ssec:unlearn_hide}

We now investigate whether unlearning leaves behind any detectable traces by examining a broad range of indicators: (i) shifts in the model’s parameter space, (ii) variations in client-side local training execution time, and (iii) differences in the communication size of model updates.

\begin{wrapfigure}{r}{0.5\textwidth}
\vspace{-20pt} 
    \centering \small
    \includegraphics[width=0.95\linewidth]{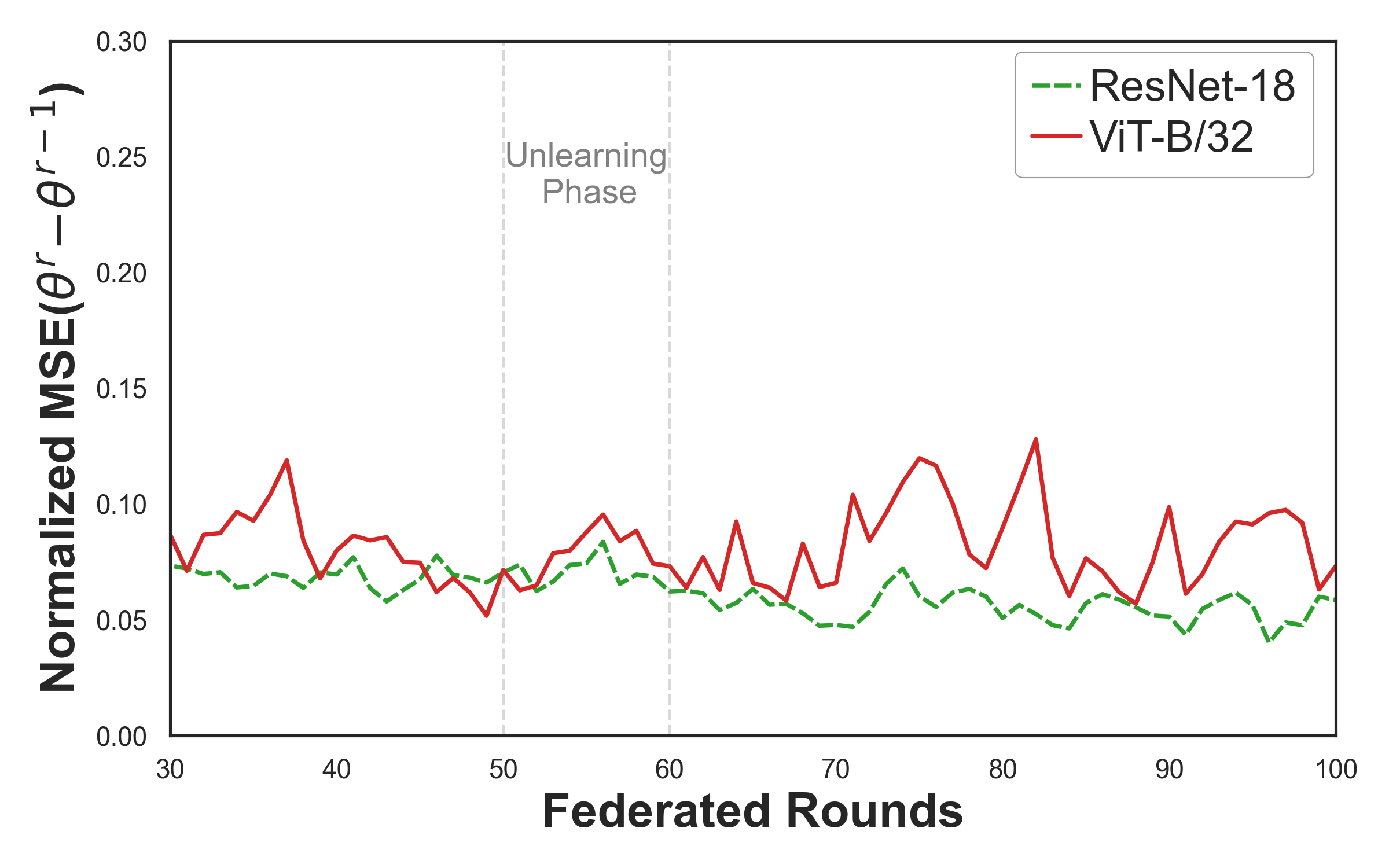}
    \caption{\emph{Model parameter variance during unlearning} with~\method. We report MSE drift between model parameters across consecutive rounds as a measure of detectability. Results shown for ResNet-18 and ViT-B/32 on CIFAR-10 for \emph{sample-wise} unlearning. Federated parameters are $N{=}10$, $N_u{=}2$, $R{=}100$, $R_u{=}50$, and $\gamma_s{=}0.1$.}
    \label{fig:param_variance}
    \vspace{-5pt} 
\end{wrapfigure}

\subsubsection{\textbf{Parameter-level detectability}}
We begin by evaluating whether unlearning induces abrupt shifts in the aggregated model parameters. To this end, we compute the Mean Squared Error (MSE) between model weights across consecutive rounds for both ResNet-18 and ViT-B/32 on CIFAR-10, normalizing by the maximum MSE — observed during early training — for comparability across models. 

As shown in Fig.~\ref{fig:param_variance}, both models exhibit stable parameter trajectories throughout the unlearning phase, with no noticeable spikes or discontinuities near the unlearning onset ($R_u{=}50$). Notably, the MSE remains within 10\% of the peak observed during training, indicating that~\method~preserves parameter-level indistinguishability and effectively hides traces of unlearning in the global model parameter-space.

\subsubsection{\textbf{Behavioral-level detectability}} 
\begin{figure*}[!t]
    \centering
    \begin{subfigure}[t]{0.48\linewidth}
        \centering\small
        \includegraphics[width=\linewidth]{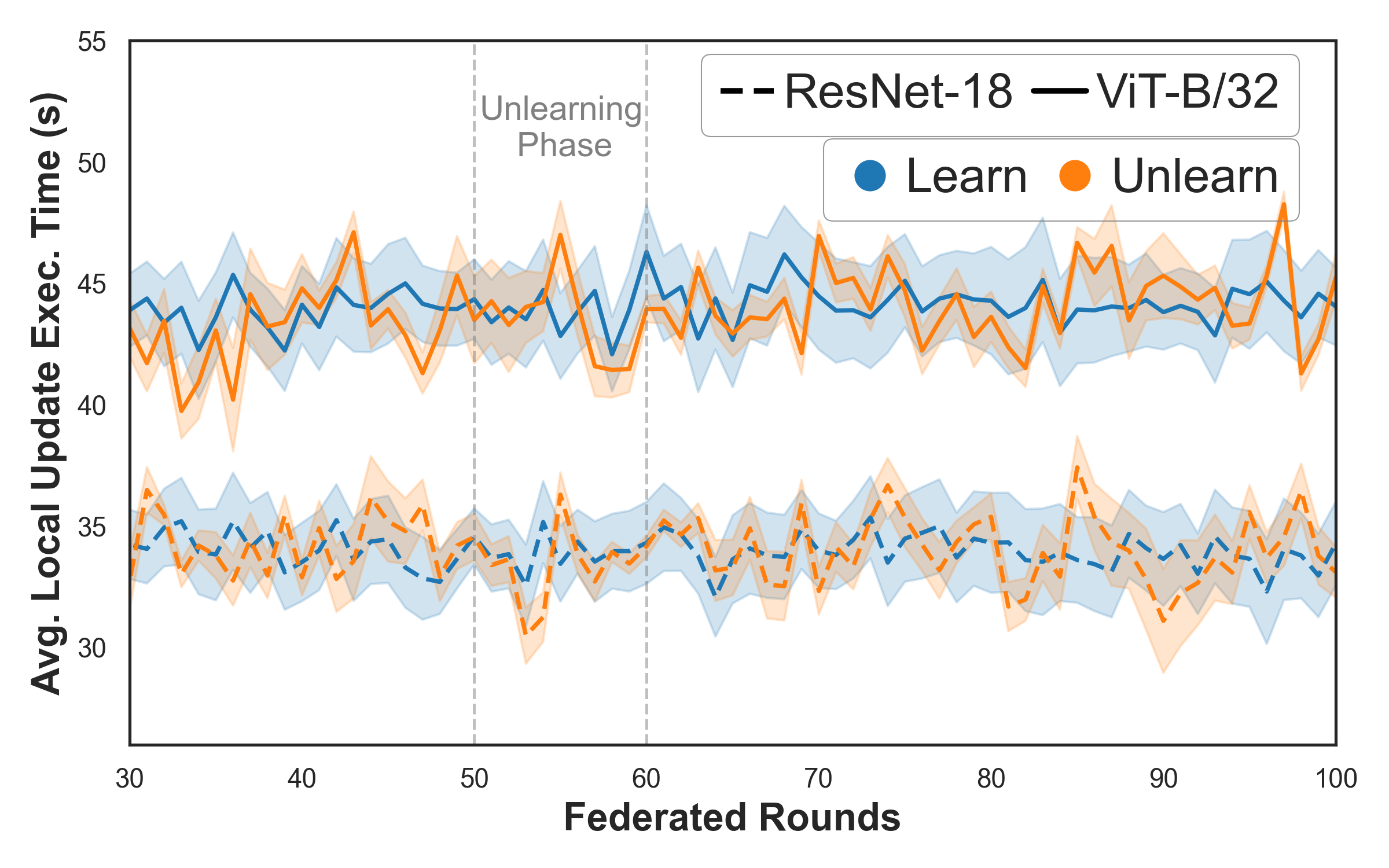}
        \caption{\emph{Local update execution time during unlearning} with~\method. We report the average per-round client-side execution time (in seconds) to assess unlearning detectability. Results shown for ResNet-18 and ViT-B/32 on CIFAR-10 for \emph{sample-wise} unlearning. Federated parameters are $N{=}10$, $N_u{=}2$, $R{=}100$, $R_u{=}50$, and $\gamma_s{=}0.1$.}
        \label{fig:exec_time}
    \end{subfigure}
    \hfill
    \begin{subfigure}[t]{0.48\linewidth}
        \centering\small
        \includegraphics[width=\linewidth]{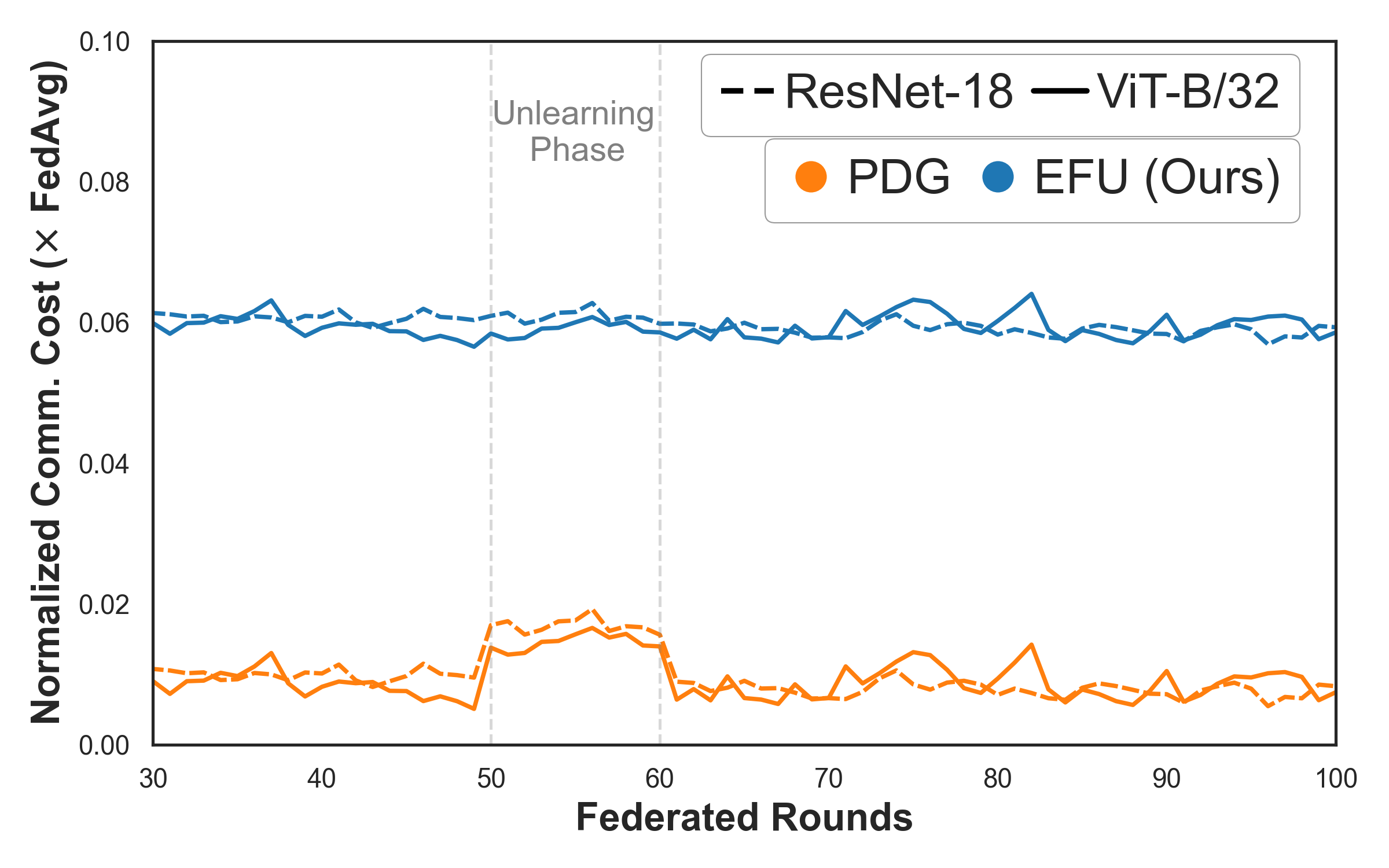}
        \caption{\emph{Model update size during unlearning} across different FU methods. We report the size of \emph{model updates} per round normalized versus standard \textit{FedAvg} as a system-level indicator of unlearning detectability. Results shown for ResNet-18 and ViT-B/32 on CIFAR-10 for \emph{sample-wise} unlearning. Federated parameters are $N{=}10$, $N_u{=}2$, $R{=}100$, $R_u{=}50$, and $\gamma_s{=}0.1$.}
        \label{fig:comm_cost}
    \end{subfigure}
    \caption{Comparison of local execution time and model update size during sample-wise unlearning using ResNet-18 and ViT-B/32 on CIFAR-10.}
    \label{fig:unlearning_overview}
\end{figure*}

Beyond conventional model-based metrics (e.g., model accuracy or its weight distribution), FL systems may inadvertently expose a client's intent to unlearn through system-level indicators — such as local computation time or communication patterns. These side channels can be exploited by adversaries to detect which clients are undergoing unlearning, undermining~\method's ``\textit{function hiding}'' guarantees. To evaluate~\method's resilience to such leakage, we monitor the average per-round local update execution time for each client using both ResNet-18 and ViT-B/32 on CIFAR-10. As shown in Fig.~\ref{fig:exec_time}, we observe no significant discrepancy between learning and unlearning rounds across either architecture, with local update execution durations remain stable and indistinguishable across the learning and unlearning phases, confirming that~\method~preserves behavioral indistinguishability and does not reveal unlearning activity through runtime side effects.

\subsubsection{\textbf{Communication-level detectability}} To evaluate whether unlearning reveals \emph{detectable client intent through communication patterns}, we analyze the per-round communication cost of client updates. Specifically, we compare~\method~to traditional gradient-based FU approaches, such as PGD~\cite{halimi2022federated}. Since model updates are rarely transmitted as raw full-weight matrices in practice, we compress the delta updates (i.e., differences from the global model) using GZip — a widely adopted mechanism in communication-efficient FL~\cite{fedzip, khalilian2023fedcode}. To facilitate direct comparison across methods and assess the communication efficiency of~\method, we normalize the compressed update sizes relative to the size of uncompressed FedAvg updates.

As shown in Fig.~\ref{fig:comm_cost}, our method maintains both a low and indistinguishable communication footprint across all rounds, with no discernible spikes during the unlearning window ($R_u{=}50{-}60$). In contrast, PGD exhibits a clear increase in communication cost during unlearning, primarily due to the transmission of dense gradient updates — unlike standard training rounds where model convergence results in sparser updates that are more effectively compressed using GZip. Importantly, since the encrypted centroids in~\method~retain a fixed size across rounds, any increase in update diversity during unlearning only affects the mappings, which operate in the integer space and are highly compressible. This contrast underscores~\method's efficiency in maintaining low and communication-indistinguishable updates, even under unlearning. Notably,~\method's communication overhead is significantly lower to standard FedAvg, echoing findings in~\cite{tsouvalas2025enccluster}, which showed that despite the introduction of FE, the communication cost is reduced by over 90\% — thanks to encrypting only a small number ($\kappa{=}64$) of floating-point centroids. 


\section{Conclusion} \label{sec:conclusion}

We presented~\method, a cryptographically enforceable FU approach that enables clients to autonomously and privately revoke the influence of their data from collaboratively trained models. Leveraging functional encryption,~\method~ensures unlearning updates are both \emph{executed} and \emph{indistinguishable} from standard training — without server trust. Our extensive evaluation show that~\method~achieves high unlearning fidelity across datasets and models, closely matching full retraining while preserving accuracy on retained data. Beyond performance,~\method~maintains indistinguishability across parameter, behavioral, and communication indicators, effectively concealing unlearning traces. Importantly, being agnostic to the unlearning strategy,~\method~can be seamlessly integrated into any FU scheme that operates by issuing tailored unlearning client updates. As such, it serves as a lightweight, \emph{drop-in} mechanism that transforms any client-side FU scheme into a secure, private, and verifiable solution. Our work highlights the need to move beyond algorithmic FU toward enforceable, privacy-preserving protocols that uphold user autonomy in real-world FL deployments.


\section*{Acknowledgement}
This work has been supported by the H2020 ECSEL EU project Distributed Artificial Intelligent System (DAIS) under grant agreement No. 101007273, and the Knowledge Foundation within the framework of INDTECH (Grant No. 20200132) and INDTECH+ Research School project (Grant No. 20220132).

\bibliographystyle{unsrt}

\clearpage \onecolumn 
\appendix

\section{Weight Clustering}\label{asec:wc}

Weight clustering is a neural network model compression technique in which similar weights are grouped into clusters using a clustering algorithm such as K-means~\cite{lloyd1982least}. This process can be executed either per layer, clustering each layer's weights independently, or across the entire model, clustering all weights collectively. Given a neural network $p_{\theta}$, parameterized by $\theta \in \mathbb{R}^d$, the objective of the weight clustering is to identify $\kappa$ distinct clusters $\mathcal{C} = \{ c_1, \ldots, c_\kappa \}$, with the aim of minimizing the following objective function:

\vspace{-3pt}
\begin{equation}\label{eqn:wc}
    \mathcal{L}_{wc}(\theta, \mathcal{Z}) = \sum_{j=1}^{\kappa} \sum_{i=1}^{d} u_{ij} \cdot || \theta_i - z_j ||^2 ~,
\end{equation}
\vspace{-3pt}

\noindent where $\mathcal{Z} = \{z_1, \dots, z_\kappa \}$ represents the set of $\kappa$ centroids, each corresponding to a cluster $c_i$. The term $||\cdot||$ denotes the Euclidean distance operator, and $u_{ij}$ is a binary indicator function that returns $1$ when weight $\theta_i$ belongs to cluster $c_j$ and $0$ otherwise. In essence, $\mathcal{L}_{wc}$ calculates the sum of squared Euclidean distances between each weight and its nearest centroid, weighted by $u_{ij}$. Upon minimizing $\mathcal{L}_{wc}$, we obtain the set of centroids $\mathcal{Z}$ and their cluster position matrix $\mathbf{P} \in \mathbb{R}^{d}$ (referred to as mappings matrix), mapping each point of $\theta$ to its corresponding cluster centroid value in $\mathcal{Z}$.

\section{Formal Definitions \& Proofs} \label{app:proofs}

\subsection{Enforceable Unlearning} 

\begin{center}
    \fbox{%
        \begin{minipage}{0.95\linewidth}
            \begin{theorem}[Decryption Enforces Complete Aggregation] \label{thm:fhe}
                Let \( \mathsf{dk}_f \) be a functional decryption key for a function \( f(\{ \theta_i^r \}_{i \in \mathcal{N}}) = \sum_{i \in \mathcal{N}} \theta_i^r \) bound to round label \( r \). Then, under IND-security of DMCFE, the server can only recover:
                \begin{equation}
                    \theta^{r+1} = \left\{ \mathsf{Dec} \left( \{ \theta_{i,\text{enc}}^{r}[j] \}_{i \in \mathcal{N}}, \mathsf{dk}_f \right) \right\}_{j=1}^d
                \end{equation}
                Decrypting any subset or altered aggregation (e.g., omitting some \( \theta_i^r \)) is computationally infeasible without a valid alternative key \( \mathsf{dk}_{f'} \), which would violate IND-security.
            \end{theorem}
        \end{minipage}%
    }
\end{center}

\begin{proof}[Sketch]
In DMCFE~\cite{chotard2018decentralized}, decryption keys are function- and label-bound: \( \mathsf{dk}_f \) only enables evaluation of the function \( f \) over ciphertexts encrypted under label \( r \). Any deviation — such as omitting terms or altering \( f \) — requires a different key \( \mathsf{dk}_{f'} \), which cannot be generated without colluding with clients. With at least one client withholds its partial decryption key for any altered function or label (e.g., the one requesting unlearning), the server cannot construct a valid \( \mathsf{dk}_{f'} \); thus, by IND-security, any attempt to decrypt with an invalid key yields no useful information, forcing the server to perform the exact aggregation over all client ciphertexts.
\end{proof}

\subsection{Update-Type Indistinguishability} 

\begin{center}
    \fbox{%
        \begin{minipage}{0.95\linewidth}
            \begin{theorem}[Update-Type Indistinguishability] \label{thm:uti}
                Let \( \mathcal{Z}_{i,\text{enc}}^r \) denote the encrypted update from client \( i \) in round \( r \), generated either through learning or unlearning. Then, for any probabilistic polynomial-time (PPT) adversary \( \mathcal{A} \), we have:
                \begin{equation}
                    \left| \Pr\left[ \mathcal{A}(\mathcal{Z}_{i,\text{enc}}^r) = \textsf{learn} \right] - \Pr\left[ \mathcal{A}(\mathcal{Z}_{i,\text{enc}}^r) = \textsf{unlearn} \right] \right|  \leq \negl(\lambda)~,
                \end{equation}
                \noindent where \( \lambda \) is the security parameter.
            \end{theorem}
        \end{minipage}%
    }
\end{center}

\begin{proof}[Sketch]
Client updates — whether produced via standard learning or unlearning — are compressed to low-rank representations \( \mathcal{Z}_i^r \), projected using \( \mathbf{P}_i^r \), and encrypted using \( \mathsf{ek}_i \) under round label \( r \). Given that ciphertexts are structurally identical and generated using an IND-secure DMCFE, no PPT adversary (including the server) can distinguish their origin beyond negligible advantage.
\end{proof}

\section{Measuring the Influence of Forgotten Data}\label{asec:mit}

In practice, exact guarantees for data removal are infeasible due to the non-convex and stochastic nature of modern ML training. Instead, machine unlearning aims to estimate how closely the behavior of the updated model after unlearning matches that of a counterfactual model trained without the forgotten data. This serves as a practical proxy for measuring the residual influence of forgotten samples. 

Formally, let \( \theta \) denote the model parameters of a neural network $p$ after applying unlearning to a forget set \( \mathcal{D}^u \) (e.g., using Eq.\,\eqref{eq:unlearn}), and let \( \theta_{\setminus \mathcal{D}^u} \) represent a counterfactual model trained identically without including \( \mathcal{D}^u \). Given a held-out validation set \( \mathcal{D}_{\text{val}} = \{(x_i, y_i)\}_{i=1}^{|\mathcal{D}_{\text{val}}|} \), we define the empirical prediction gap between the two models, \( \varepsilon \), as:

\vspace{-3pt}
\begin{equation}
    \varepsilon = \frac{1}{|\mathcal{D}_{\text{val}}|} \sum_{(x_i, y_i) \in \mathcal{D}_{\text{val}}} \left\| p_{\theta}(x_i) - p_{\theta_{\setminus \mathcal{D}^u}}(x_i) \right\|,
\end{equation}
\vspace{-3pt}

\noindent where \( p_{\theta}(x_i) \) and \( p_{\theta_{\setminus \mathcal{D}^u}}(x_i) \) are the output distributions of the model after unlearning and the counterfactual model (trained without \( \mathcal{D}^u \)), respectively. Intuitively, the scalar \( \varepsilon \) quantifies the average per-sample prediction deviation between the two models on a held-out validation set \( \mathcal{D}_{\text{val}} \), where a lower \( \varepsilon \) indicates closer behavioral alignment with the counterfactual model, thereby serving as an empirical proxy for the extent to which the influence of \( \mathcal{D}^u \) has been removed.


\end{document}